\documentclass[preprint2]{aastex}

\usepackage{graphicx,rotate}
\usepackage{natbib,lscape}
\usepackage{latexsym,amssymb,verbatim}

\shorttitle{extragalactic CS survey}
\shortauthors{Bayet et al.}

\begin{document}


\title{Extragalactic CS survey}


\author{E. Bayet\altaffilmark{1}, R. Aladro\altaffilmark{2},
S. Mart\'in\altaffilmark{3}, S. Viti\altaffilmark{1}, and J.
Mart\'in-Pintado\altaffilmark{2}}

\email{eb@star.ucl.ac.uk}


\altaffiltext{1}{Department of Physics and Astronomy, University
College London, Gower Street, London WC1E 6BT, UK.}
\altaffiltext{2}{Departamento de Astrofisica Molecular e
Infrarroja - Instituto de Estructura de la Materia-CSIC, C Serrano
121, E-28006 Madrid, Spain} \altaffiltext{3}{Harvard Smithsonian
Center for Astrophysics, 60 Garden Street, Cambridge, MA 02138,
USA}


\begin{abstract}

We present a coherent and homogeneous multi-line study of the CS
molecule in nearby (D$<$10Mpc) galaxies. We include, from the
literature, all the available observations from the $J=1-0$ to the
$J=7-6$ transitions towards NGC~253, NGC~1068, IC~342,
Henize~2-10, M~82, the Antennae Galaxies and M~83. We have, for
the first time, detected the CS(7-6) line in NGC~253, M~82 (both
in the North-East and South-West molecular lobes), NGC 4038, M~83
and tentatively in NGC~1068, IC~342 and Henize~2-10. We use the CS
molecule as a tracer of the densest gas component of the ISM in
extragalactic star-forming regions, following previous theoretical
and observational studies by Bayet et al. (2008a,b and 2009). In
this first paper out of a series, we analyze the CS data sample
under both Local Thermodynamical Equilibrium (LTE) and non-LTE
(Large Velocity Gradient-LVG) approximations. We show that except
for M~83 and Overlap (a shifted gas-rich position from the nucleus
NGC~4039 in the Antennae Galaxies), the observations in NGC~253,
IC~342, M~82-NE, M~82-SW and NGC~4038 are not well reproduced by a
single set of gas component properties and that, at least, two gas
components are required. For each gas component, we provide
estimates of the corresponding kinetic temperature, total CS
column density and gas density.

\end{abstract}


\keywords{Galaxies: nuclei-ISM-individual: NGC~253, NGC~1068,
IC~342, Henize~2-10, M~82, The Antennae, M~83 -- millimeter --
ISM: molecules -- methods: data analysis}



%-------------------------------------------------------------------

\section{Introduction}\label{sec:intro}

Detection of star-forming gas is one of the most direct ways to
measure the star formation rate and activity in a galaxy, allowing
us to significantly improve our understanding of galaxy formation
and evolution. Star-forming regions of very dense gas
(n(H$_{2}$)$> 10^{5}$cm$^{-3}$) are needed to maintain star
formation activity, even in hostile environments associated with
young massive stars. In these environments, star-forming regions
are able to resist the disruptive forces (winds or radiation) from
nearby newly formed stars longer than the local gas in the local
interstellar medium \citep{Klei83, LaRo83}. Determining the
physical conditions of the very dense gas in which massive stars
form is, thus, essential. In this paper, we study such dense gas
and estimate its properties over a large range of nearby galaxy
types.

Following theoretical studies by \citet{Baye08a} and the first
detections of extragalactic very dense gas presented in
\citet{Maue89a, Maue89b, Walk90, Mart05, Mart06b, Mart06a,
Baye08b, Grev09}, we have carried out multi-line observations of
the CS molecule in nearby (D$<$10Mpc) extragalactic environments,
enhancing significatively the current data set of extragalactic CS
observations. Indeed, CS lines have not been observed so far but
in few brightest nearby nuclei such as NGC~253, IC~342 and M~82
and mainly in their lower-J rotational levels (see references
above). Hence, the physical properties of such galaxies namely the
kinetic temperature, gas density, etc were estimated using only a
small sample of CS lines. Gas traced by higher-J of species such
as CS has not be characterized so far. In addition, even for the
brightest sources (e.g. NGC~253 and M~82), no study of the various
velocity components nor various positions in the same galaxy has
been performed so far in a systematic way. In this paper, first of
a series, we thus aim at investigating in much more details the
very dense gas properties in many extragalactic sources as traced
by the CS molecule.

Sulphur-bearing species are shown to be particularly enhanced
during massive star formation, while species such as HCN, although
a useful dense gas tracer, may not be tracing the sites where star
formation occurs \citep{Lint06}. Recently, \cite{Mart05} showed
that sulfur emission in the nuclear region of the nearby starburst
galaxy NGC~253 is very strong. Amongst sulfur-bearing species, the
CS molecule appears as one of the best tracers of dense gas,
especially its high-J rotational transitions such as the CS
$J=5-4$ line with an excitation threshold higher than
10$^{4}$-10$^{5}$ cm$^{-3}$ \citep{Bron96} and the CS $J=7-6$ line
with a critical density of n$_{crit} \sim 2\times 10^{7}$
cm$^{-3}$ \citep{Plum92}.

Selection of nearby sources where we have proposed to observe the
CS lines was made through a comparison with past and recent CS
observations (see references above). Sources where no CS
detections were available from the literature (e.g. Henize 2-10)
were chosen on the basis of the detections of the CO(6-5) or
CO(7-6) lines since these lines are also good tracers of a
relatively dense and quite warm gas \citep{Baye04, Baye06}. The
source selection also aimed at presenting various nearby galaxy
types where CS line emission can be studied and compared. This
sample is far from being unbiased but it gathers so far the most
complete CS data sample ever obtained in extragalactic sources.

Thus, the centers of NGC~253, IC~342, Henize~2-10, the two
molecular lobes of M~82 (North-East and South-West positions),
three positions in the Antennae Galaxies (the two nuclei: NGC~4038
and NGC~4039, and a shifted position from NGC~4039 called Overlap)
and the center of M~83 were selected, covering a large range of
galaxy types and star-formation activities (starburst, irregular,
merging galaxies, more quiet star-forming galaxies, etc). We have
also included in our source sample the center of NGC~1068
(AGN-dominated galaxy) where HCN has been detected
\citep{Plan91,Krip08}.

We present our observations, data reduction and results in Sect.
\ref{sec:obs}. In Sect. \ref{sec:mol} we analyze the data set both
under Local Thermodynamical Equilibrium (LTE) and non-LTE (Large
Velocity Gradient-LVG) approximations (see Subsects.
\ref{subsec:lte} and \ref{subsec:lvg}). In Sects. \ref{sec:disc}
and \ref{sec:con} we discuss our findings and conclude,
respectively.

\section{Observations and Results}\label{sec:obs}

\subsection{Observational parameters}

\begin{figure}
\includegraphics[height=3cm]{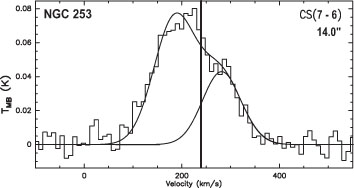}
\caption{CS(7-6) spectrum in the center of NGC~253. The angular
resolution is shown in the upper right corner. The thick vertical
black line symbolizes the V$_{LSR}$ of the source. For NGC~253, we
have applied a two-components fit (thin black lines) after having
smoothed the entire signal to a velocity resolution of $\approx
10$ kms$^{-1}$. The fit, on the left is for the emission from
NGC~253-1 while on the right, it is for the NGC~253-2
emission.}\label{fig:1}
\end{figure}

\begin{figure}
\includegraphics[height=3cm]{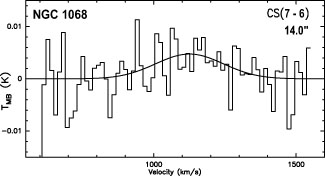}
\caption{Marginal detection of the CS(7-6) line in the center of NGC~1068
($\Delta$v$=13.7$kms$^{-1}$).}\label{fig:2}
\end{figure}

\begin{figure}
\includegraphics[height=3cm]{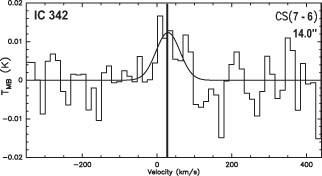}
\caption{Marginal detection of the CS(7-6) line in the center of IC~342
($\Delta$v$=13.7$kms$^{-1}$).}\label{fig:3}
\end{figure}

\begin{figure}
\includegraphics[height=3cm]{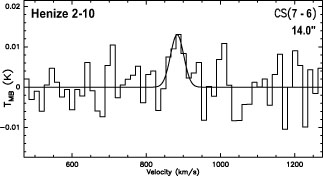}
\caption{Marginal detection of the CS(7-6) line in the center of Henize~2-10
($\Delta$v$=13.7$kms$^{-1}$).}\label{fig:4}
\end{figure}

\begin{figure}
\includegraphics[height=14cm]{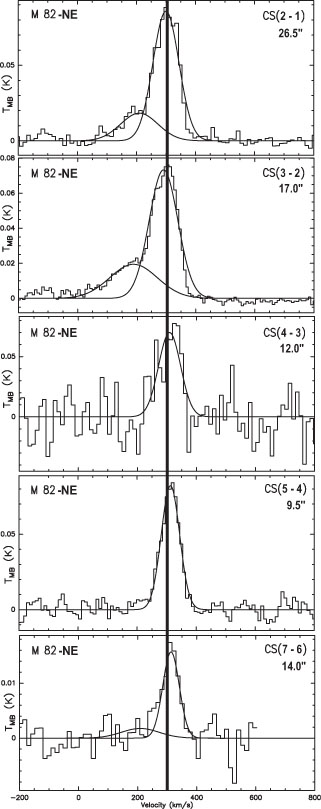}
\caption{CS spectra in the NE molecular lobe of M~82.
The velocity resolution from top to bottom is
$\Delta$v$=12.2$kms$^{-1}$, 8.1kms$^{-1}$, 24.5kms$^{-1}$,
9.7kms$^{-1}$ and 13.7kms$^{-1}$, respectively.}\label{fig:5}
\end{figure}

\begin{figure}
\includegraphics[height=9cm]{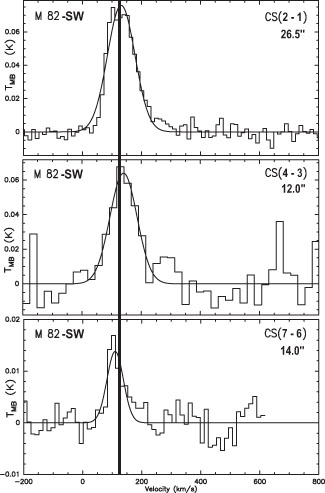}
\caption{CS spectra in the SW molecular lobe of M~82. The velocity
resolution from top to bottom is $\Delta$v$=12.2$kms$^{-1}$,
24.5kms$^{-1}$ and 13.7kms$^{-1}$, respectively.}\label{fig:6}
\end{figure}

\begin{figure}
\includegraphics[height=18cm]{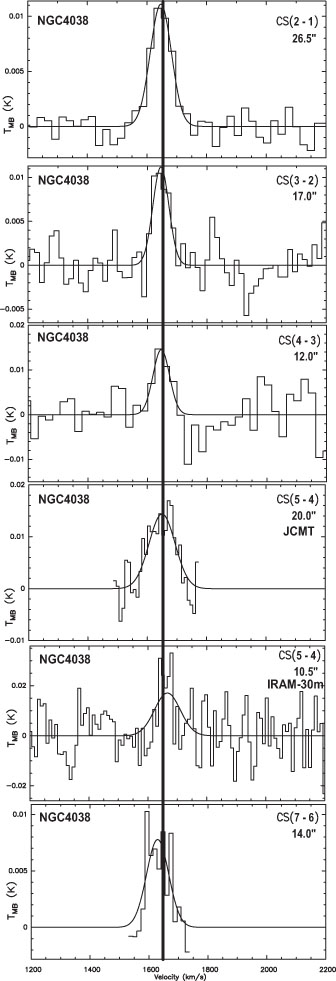}
\caption{CS spectra in NGC~4038 (one of the nuclei in The Antennae
Galaxies). The velocity
resolution from top to bottom is $\Delta$v$=24.5$kms$^{-1}$,
16.3kms$^{-1}$, 24.5kms$^{-1}$, 10.0kms$^{-1}$, 10.0kms$^{-1}$ and
13.7 kms$^{-1}$, respectively.}\label{fig:7}
\end{figure}

\begin{figure}
\includegraphics[height=12cm]{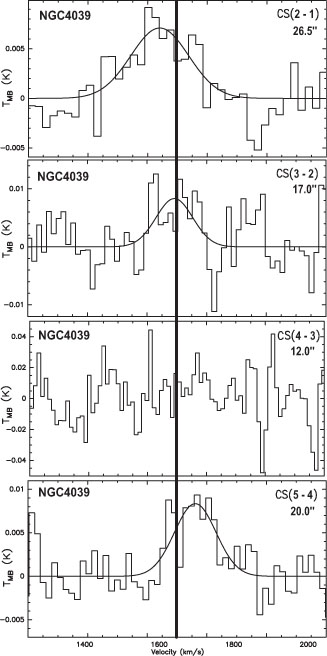}
\caption{CS spectra in NGC~4039 (one of the nuclei in The Antennae
Galaxies). The velocity resolution from top to bottom is
$\Delta$v$=24.5$kms$^{-1}$, 16.3kms$^{-1}$, 12.2kms$^{-1}$ and
19.1kms$^{-1}$, respectively.}\label{fig:8}
\end{figure}

\begin{figure}
\includegraphics[height=15cm]{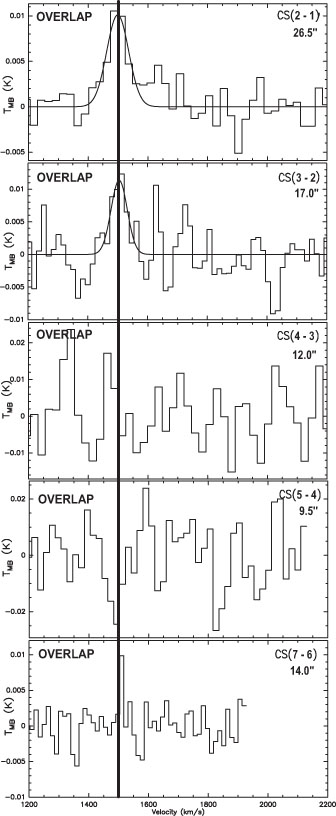}
\caption{CS spectra in the Overlap position (The Antennae
Galaxies). Due to poor weather conditions, no detection is seen
neither in the CS(4-3), nor the CS(5-4) nor the CS(7-6) lines. The
velocity resolution from top to bottom is
$\Delta$v$=24.5$kms$^{-1}$, 16.3kms$^{-1}$, 24.5kms$^{-1}$,
19.6kms$^{-1}$ and 13.7kms$^{-1}$, respectively.}\label{fig:9}
\end{figure}

\begin{figure}
\includegraphics[height=3cm]{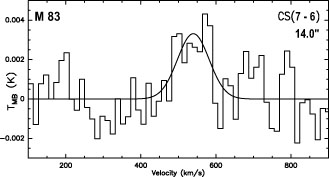}
\caption{CS(7-6) spectrum in the center of M~83
($\Delta$v$=13.7$kms$^{-1}$).}\label{fig:10}
\end{figure}

The observations have been performed using the IRAM-30m telescope
for the CS(2-1), CS(3-2) and CS(4-3) lines ($\nu=$ 97.980 GHz,
146.969 GHz and 195.954 GHz, respectively) and the James Clerk
Maxwell Telescope (JCMT) for the detection of the CS(7-6)
transition ($\nu=$ 342.883 GHz). Every two-three hours, when using
both telescopes, the pointing, focus and calibration were
performed carefully on planets (Mars and Jupiter) and on evolved
stars. The pointing error was estimated to be $\leq$ 3$''$ in both
cases. We have observed the CS(5-4) line with both telescopes but
we have not averaged the signals because the weather conditions
were significantly different. Between the two telescope
observational sessions, for all the sources but NGC~4038, we have
kept the less noisy spectrum (i.e. the spectrum showing the
highest S/N ratio). These (reduced) spectra are shown in Figs.
\ref{fig:1} to \ref{fig:10} and their Gaussian fitting parameters
are seen in Table \ref{tab:obs}. Unfortunately, for NGC~4038, the
reduced spectra obtained with the IRAM-30m and with the JCMT are
equally noisy (i.e. they have similar S/N ratio) and thus we
present both spectra in Fig.\ref{fig:7}.

The IRAM-30m telescope observations were carried out in June 2008
under various weather conditions ($\approx$ 1.5-6 mm of water
vapor). Scans were taken using the wobbling secondary with a throw
of 120$''$ for the two positions in M~82 (NE and SW components)
while for the Antennae positions (NGC~4038, NGC~4039 and Overlap)
we have used the position switching mode with a larger throw of
1$'$ (reference positions are indicated in columns 12 and 13 of
Table \ref{tab:obs}). The 100 GHz, 150 GHz, 230 GHz and 270 GHz
receivers (A/B/C/D) were used under several configurations,
optimizing the allocated time. The 1 MHz and 4 MHz backends were
used providing velocity resolution between 1.2 kms$^{-1}$ (at 244
GHz) and 3.0 kms$^{-1}$ (at 98 GHz), and, 4.9 kms$^{-1}$ (at 244
GHz) and 12.3 kms$^{-1}$ (at 98 GHz), respectively. At the
observed frequencies of 97.980 GHz, 146.969 GHz, 195.954 GHz and
244.936 GHz, the IRAM-30m has a HPBW of $\approx$ 25$''$, 17$''$,
13$''$ and 10$''$ with main beam efficiencies of 0.75, 0.69, 0.65,
0.52\footnote{See IRAM-30m website: http://www.iram.es/IRAMES/
mainWiki/AbcdforAstronomers}, respectively. The system temperature
ranged from $\approx$ 200 K to $\approx$ 1800 K, depending on the
frequency and the weather conditions. The data reduction was
performed using the CLASS program of the GILDAS package developed
at IRAM.

We have used the JCMT in March 2007, February 2008, July 2008 and
February 2009 under medium weather conditions ($\tau_{225}
\approx$ 0.15). We have either used a beam switch mode with a
throw of 180$''$ or a position switch mode with hard off-source
reference, depending on the source. We have used the RxA3 and the
16-pixel camera HARP receivers for the detection of the CS(5-4)
and the CS(7-6) transitions, respectively. The ACSIS digital
autocorrelation spectrometer with a bandwidth of 1000MHz was used
at both frequencies and in all the sources because the lines were
expected to be broad (see Figs. \ref{fig:1} to \ref{fig:10}). The
HPBW and the main beam efficiencies of the JCMT at $\nu=$ 244 GHz
and $\nu=$ 343 GHz are 20$''$, 14$''$ and 0.69, 0.63\footnote{See
JCMT website: http://www.jach.hawaii.edu/JCMT/ instruments/},
respectively (see Table \ref{tab:obs}). The system temperatures
ranged between 200 K and 400 K. The data pre-reduction was done
using Starlink softwares (KAPPA, SMURF, and STLCONVERT packages)
and subsequently translated to CLASS format for final reduction.

All the spectra obtained are shown in Figs. \ref{fig:1} to
\ref{fig:10} while the Gaussian fit parameters are listed in Table
\ref{tab:obs}. In this Table, we have also included detections of
CS lines from the literature (see references mentioned in the last
column of the table). Hence this table contains the \textit{most
complete compilation of CS transitions observed in external
galaxies}.

\subsection{Results}

For each source, we find a general good agreement between the
Gaussian fits and the observations (see Figs. \ref{fig:1} to
\ref{fig:10}). Table \ref{tab:obs} also shows that the velocity
positions of the lines and the line widths of the CS transitions
are consistent among each other and with previous data from the
literature.

For some sources, some lines of CS are only marginally detected.
For instance, the CS(7-6) line in NGC~1068 (Fig. \ref{fig:2}) and
in IC~342 (Fig. \ref{fig:3}) have both a signal-to-noise ratio
lower than 3 and thus can not been considered as detections. We
have also obtained for NGC~4039 a hint of a marginal detection but
due to the uncertainty on the velocity position of the line, we
have not shown the spectrum. Except for NGC~4039, in all the
marginal cases presented here, the fitted line width and velocity
position match those expected either from lower-J CS lines (when
available) or in the case where no other CS lines are available,
from other molecular lines such as CO as in the case of the
CS(7-6) line in Henize~2-10. For this source, we have derived an
upper limit of the source-averaged CS column density of
N(CS)$=7.3\times 10^{12}$cm$^{-2}$ from the CS(7-6) marginal
detection, using a source size of 13$''$ (see \citealt{Meie01b,
Baye04}). Due to the fact that there is only one CS observation so
far (see Fig. \ref{fig:4}), we have excluded Henize~2-10 from the
data analysis presented in Sect. \ref{sec:mol}.

For NGC~253, based on the literature, it is expected that some CS
detections show a double peak emissions corresponding to two gas
components (see \citealt{Mart05} and Fig \ref{fig:1}). Thus, in
NGC~253, we have called NGC~253-1, the gas component whose
line emission is located at a velocity position of about
180-200kms$^{-1}$ while NGC~253-2 corresponds to the higher
velocity gas component located at about 280-300kms$^{-1}$. A
two-components Gaussian fit has thus been applied. We have listed
the fit parameters obtained in Table \ref{tab:obs}.

In some of the M~82-NE CS spectra (see Fig. \ref{fig:5}), one also
notice a double peak emission. However the nature of this emission
differs significantly from the one observed in NGC~253. Indeed,
after having compared Fig. \ref{fig:5} with the data from
\citet{Baye08b}, we have concluded that the weaker peak (seen on
the left side of the spectra) corresponds to a contamination from
the nucleus (partially included into the telescope beam). We have
thus excluded such emission from the analysis performed in
Subsects. \ref{subsec:lte} and \ref{subsec:lvg}.

\section{Molecular excitation}\label{sec:mol}

\subsection{LTE analysis}\label{subsec:lte}

\begin{figure}
\includegraphics[height=10.5cm]{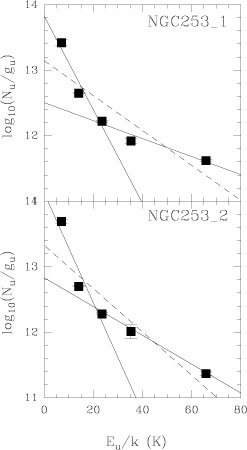}
\caption{CS rotational diagrams of NGC~253-1 (top plot) and
NGC~253-2 (bottom plot). The linear regression using a
single-component is represented by dashed black lines while the
two-components regression is represented by solid black lines.
Detections are represented by black squares symbols with error
bars. These error bars actually correspond to those of the
integrated intensities (order of 10-20\%). When a CS line is
marginally detected (upper limit), we represent the corresponding
datum with an open white square (see also Figs. \ref{fig:12},
\ref{fig:13} and \ref{fig:15}).}\label{fig:11}
\end{figure}

\begin{figure}
\includegraphics[height=5.5cm]{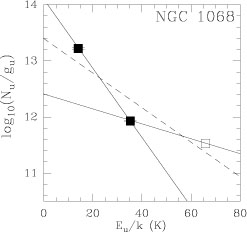}
\caption{CS rotational diagram of the center of
NGC~1068.}\label{fig:12}
\end{figure}

\begin{figure}
\includegraphics[height=5.5cm]{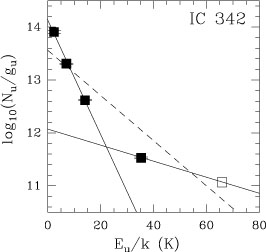}
\caption{CS rotational diagram of the center of
IC~342.}\label{fig:13}
\end{figure}

\begin{figure}
\includegraphics[height=10cm]{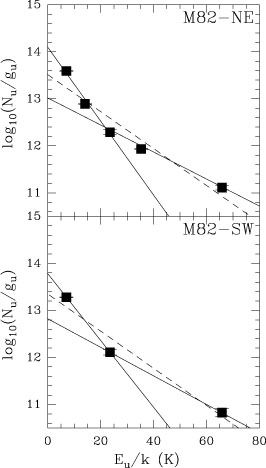}
\caption{CS rotational diagrams for the M~82-NE (top plot) and the
M~82-SW (bottom plot) molecular lobes. }\label{fig:14}
\end{figure}

\begin{figure}
\includegraphics[height=14cm]{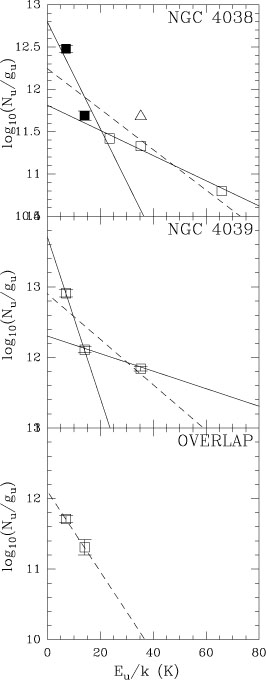}
\caption{CS rotational diagrams of the three positions observed in
the Antennae: NGC~4038 (top plot), NGC~4039 (middle plot) and
Overlap (bottom plot). For the NGC~4038 CS(5-4) line, we have used
a white open triangle and a white open square for representing the
IRAM-30m and the JCMT observations, respectively.}\label{fig:15}
\end{figure}

\begin{figure}
\includegraphics[height=5cm]{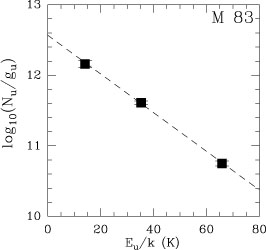}
\caption{CS rotational diagram of the center of M~83.
}\label{fig:16}
\end{figure}

To obtain an estimation of the kinetic temperature and the total
CS column density in the various extragalactic star-forming
regions we survey here, we have used first the rotational diagram
method, corrected for beam dilution effects and assuming optically
thin emission (antennae temperature proportional to the column
density in the upper level of the observed transition). The beam
dilution effect has been removed assuming a source size of 20$''$,
20$''$, 4$''$, 6$''$, 6$''$, 8.5$''$, 7$''$, 5.5$''$, 20$''$ and
5$''$ for NGC~253-1, NGC~253-2, NGC~1068, IC~342, M82-NE, M~82-SW,
NGC~4038, NGC~4039, Overlap and M~83, respectively. These values
have been selected from the available maps of CS published in the
literature or from maps of other tracers of dense gas. When
neither was available, we have used, by default, interferometric
CO maps. Namely, for the NGC~253 source size, we refer to the
CS(2-1) map of \citet{Peng96}, in agreement with the values used
for SO$_{2}$, NS and NO emission (tracers of relatively dense gas)
presented in \citet{Mart03}. For NGC~1068, we have used the HCN
maps from \citet{Helf95} and \citet{Krip08} for deriving the CS
emitting source size. For IC 342, we base our estimation on the
HCN interferometric map of \citet{Schi08} whereas for M~82, we
used SiO and HCO interferometric maps from \citet{Garc01,Garc02},
respectively. For the three positions in the Antennae Galaxies, we
have used the interferometric CO map from \citet{Wils00} as an
upper limit of the CS emitting source size while for M~83, we have
used the recent HCN interferometric map of \citet{Mura09}. In
Figs. \ref{fig:11} to \ref{fig:16}, we show the resulting
rotational diagrams for all the sources.

The rotational diagram approach (see \citealt{Gold99} and appendix
in \citealt{Turn91, Gira02, Mart05}) assumes the gas in Local
Thermodynamical Equilibrium (LTE). If correct, then the data are
well fitted by a linear regression using a single-component. In
fact, assuming LTE in rotational diagrams implies that the
rotational temperature is equal to the excitation temperatures of
all the observed transitions. As the gas may not be thermalized,
the derived rotational temperature is thus a lower limit to the
kinetic temperature.

The LTE solutions are shown in Figs. \ref{fig:11} to \ref{fig:16}
by dashed lines (single-component fit)\footnote{The fits have been
performed by the xmgrace software.}. As clearly shown by the
figures, most of the data are not well fitted by a
single-component fit. This is not surprising since it is expected
that the gas contained in the nucleus of galaxies such as NGC~253,
IC~342, etc is not in thermal equilibrium. Thus, to better
reproduce the data, we have adopted a very simple approach
assuming that, when the single-component fit is not applicable, at
least, two gas components exist in the observed regions: one
responsible for the low-J CS emission and one for the high-J CS
line emissions. The fits performed with two components are shown
by solid lines in Figs. \ref{fig:11} to \ref{fig:15}. Note however 
that when a source shows two clear detections and an upper limit, 
our approach is crude. Whereas we
simply fitted the rotation diagrams with two temperature
components, it is clear that there is a continuum change of the
excitation temperature.

The exceptions to this approach are M~83 and the Overlap position.
Indeed, for these sources, the CS data appear to be well fitted by
a single-component fit. This means that, in these sources, the gas
might be more homogeneously distributed than in other sources.
This feature has already been observed in M~83 with the CO data
\citep{Baye06}. \citet{Baye06} indeed showed that the same gas
component well reproduce the CO(3-2), CO(4-3) and CO(6-5) lines in
the center of M~83, the requirement for another (colder) gas
component appearing only when considering lower-J CO lines. The
current rotational diagram for M~83 and Overlap contains only
three and two CS detections, respectively. It may be that
two-components fit may be necessary once the CS(1-0), CS(2-1),
CS(4-3) lines will be observed in M~83 and when the CS(1-0),
CS(4-3), CS(5-4) and CS(7-6) transitions will be added to the
rotational diagram of Overlap.

In Figs. \ref{fig:11} to \ref{fig:16}, we have used different
symbols to represent marginal detections (open white squares) and
detections (black filled squares). For NGC~4038, two CS(5-4)
(marginal) detections exist. We have shown the IRAM-30m one with
an open white triangle and the JCMT with an open white square
since it appears in better agreement with the rest of the data and
shows a higher signal-to-noise ratio (see Fig. \ref{fig:7} and
Table \ref{tab:obs}).

The rotational temperatures we have obtained for each source are
listed in Table \ref{tab:T_rot}. The typical uncertainties on the
rotational temperatures are of 5\%-6\%. We have also derived the
source-averaged total column densities N(CS) for each component in
each source (see Table \ref{tab:T_rot}). For determining such
values, we have used the partition function from the CDMS
website\footnote{See
http://www.astro.uni-koeln.de/site/vorhersagen/catalog/
partition\_function.html}, extrapolating their values to the range
of rotational temperatures listed in Table \ref{tab:T_rot} (linear
interpolation applied).

Below, we compare our results with those available in the literature.

\begin{itemize}
\item[-] \textbf{NGC~253}: As expected, the CS column densities
for both NGC~253-1 and NGC~253-2 sources (low- and high-velocity
component of NGC~253) are in agreement with the values derived by
\citet{Mart05}. Indeed, for the NGC~253-1 source the CS column
density we have computed for the single-component fit is in
agreement within a factor of 0.9. For the NGC~253-2 source, the CS
column density for the single-component fit is in agreement within
a factor of 1.8. The differences between these two studies
originate clearly from the presence of the new CS(7-6)
observations. This transition reveals or confirms in most of the
cases the need of at least one more component to better
reproduce the data.\\
\item[-] \textbf{IC~342}: \citet{Maue89b} have also observed in
 various CS transitions the center of the galaxy IC 342. They estimated
 the column density to be N(CS)=$1.5 \times 10^{14}$cm$^{-2}$, in
 agreement (with a factor of 0.5) with our estimate
 (single-component fit).\\
\item[-] \textbf{Molecular lobes in M~82}: M~82-NE and M~82-SW
 lobes have never been observed previously in any CS lines.
 However, \citet{Baye08b} showed that in the center of M~82, the CS
 column density is estimated to be N(CS)=$0.13-6.7 \times
 10^{14}$cm$^{-2}$. This is of the same order of magnitude of the
 values presented in Table \ref{tab:T_rot} (single-component
 fit) for the lobes positions.\\
\item[-] \textbf{The Antennae Galaxy}: Similarly to the lobes in
 M~82, no CS detection has been published so far neither in
 NGC~4039 nor in the Overlap position in the Antennae Galaxies.
 Only for NGC~4038 \citet{Baye08b} have obtained N(CS)=$1.8 \times
 10^{13}$cm$^{-2}$ (from the single CS(5-4) observation). This
 estimate is in agreement with the values listed in Table
 \ref{tab:T_rot} (single-component fit).\\
\item[-] \textbf{M~83}: The CS column density computed when adding
 the CS(7-6) line to the data available from the literature is in
 agreement with that derived by \citet{Maue89b} and \citet{Mart09}
 when using their respective single transition of CS.\\
\end{itemize}

In the Antennae Galaxies case, using CO data for representing the
more compact (CS) emitting source size is misleading and can cause
large uncertainties on the results as shown by \citet{Helf95} for
NGC~1068. Taking NGC~4038 as a representative example, we have
thus investigated the influence of the source size choice on the
results (LTE and non-LTE models, see Subsect. \ref{subsec:lvg}) in
the following way:

\begin{itemize}
\item[1.] We have compared the physical parameters of column
 density and temperature derived assuming a source extent of 7
 $''$, with those obtained by assuming sizes of $75\%$, $50\%$ and
 $25\%$ smaller (i.e. 5.25 $''$, 3.5 $''$ and 1.75 $''$) similar to
 all transitions. We have rebuilt consistently
 the rotational diagrams and re-calculated in the same way as
 previously described, the rotational temperature and the total CS
 column densities values. When these values were compared to those obtained in
 the case of a 7 $''$ source size, we have obtained rotational temperatures
 difference less than 7\% and column densities differences within a
 factor of $<$ 14.8 (for the smallest size considered), whatever the
 number of component considered
 (see Table \ref{tab:T_rot}).\\

\item[2.] We compared the result of a common 7 $''$ source size
 for all transitions with the case of a gradually smaller size for
 the higher transitions. Starting with a source size of 7 $''$ for
 the CS(2-1) transition, we assumed a 20\% smaller size for the 3-2
 and the 4-3 (i.e. 5.6$''$) and an additionally 20\% decrease in
 the size for the 5-4 and the 7-6 (i.e. $4.5''$). When comparing
 these cases with the case of a 7$''$ source size applied
 homogeneously to all the CS lines, it resulted a 15\%-30\%
 difference in rotational temperatures and less than a factor of 2
 in column density.\\
\end{itemize}

As shown in both cases (see Table \ref{tab:T_rot}), though the
column densities might be significantly affected by the assumed
source extent, both the derived rotational temperatures and
therefore the different temperature components described in this
paper will be basically unaffected.

In all the cases dealt with above, the CS line ratios used for
constraining the LVG model do change by no more than a factor of
1.5 (for the CS(7-6)/CS(4-3) line ratio). Considering the
degeneracy of the LVG models, as noted in Sect. \ref{subsec:lvg},
we thus believe that these variations of line ratios are
negligible and that the results presented in Table \ref{tab:LVG}
still provide qualitative estimates of the properties of the gas
emitting the CS lines.

\subsection{non-LTE models}\label{subsec:lvg}

It is well-known that rotational diagrams give only approximate
estimates of the kinetic temperature and the column density. To
get more accurate values, we have ran Large Velocity Gradients
(LVG) models \citep{Gold74, DeJo75}. These models are described in
various papers and the version we use in this study is the one
presented in \citet{Baye04, Baye06, Baye08b}. The LVG model runs
with three free parameters: the gas density n(H$_{2}$), the
kinetic temperature (T$_{k}$) and the CS column density divided by
the line width (N($^{12}$CS)$/ \Delta v$). Using as model
constraints the integrated line intensities ratios computed from
Table \ref{tab:T_rot}, and corrected for beam dilution effect (as
done in Subsect. \ref{subsec:lte}), we have investigated the
following range of LVG input parameters : 5K $<$T$_{K}<$ 150K, $1
\times 10^{11}$cm$^{-2}$(kms$^{-1})^{-1} <$N($^{12}$CS)$/ \Delta
v$\footnote{We have assumed for all the sources and line profiles
a $\Delta v \approx 100$kms$^{-1}$}$< 1 \times
10^{16}$cm$^{-2}$(kms$^{-1})^{-1}$ and $1 \times 10^{4}$cm$^{-3}
<$n(H$_{2}$)$< 1 \times 10^{7}$cm$^{-3}$.

For determining the set of parameters (T$_{K}$, n(H$_{2}$),
N($^{12}$CS)) able to reproduce best the observations, we have
constrained the predicted CS line intensity ratios with the
observed values via a reduced $\chi^{2}$ method as performed in
\citet{Baye04, Baye06, Baye08b}, including 20\% uncertainties on 
the observed integrated line intensities. Except for M~83 and Overlap, we
have separated the models results into two: the low-temperature
component constrained by, for example, line intensity ratios such
as CS(3-2)/CS(2-1) or CS(4-3)/CS(2-1), and the high-temperature
component characterized by, for example, line intensity ratios
such as CS(5-4)/CS(4-3) or CS(7-6)/CS(4-3). The choice of the CS
line intensity ratios used for calculating the $\chi^{2}$ depends
on the source and the available data set for CS (see Table
\ref{tab:obs}). For M~83 and Overlap, since the CS data are well
arranged alongside a line in the rotational diagram, only a
single-temperature component model has been investigated using the
available CS detections. The physical properties which reproduce
best the observations are summarized in Table \ref{tab:LVG}.

Unfortunately, as already shown by e.g. \citet{Baye04, Baye06} and
\citet{Mart05, Mart06b}, there is no unique solution for
reproducing the observational line ratios of each gas component,
in each galaxy. Indeed, it is well-known that degeneracies in the
physical parameters predictions exist in LVG models. To remove
this degeneracy and obtain more accurate estimations of the
physical parameters, one should correlate LVG models solutions
obtained for various molecular species. However, this methodology
is only effective if all species are emitted from the same gas
phase (see e.g. \citealt{Mart05, Mart06b}). For most of the
sources observed here, no other tracers of very dense gas
(n(H$_{2}$) $\geq 10^{5-6}$) has been detected so far. We thus did
not cross correlate our LVG results with those from other
molecular emission.

\section{Discussion}\label{sec:disc}

We find that CS is indeed tracing very dense gas. For most
sources, low-J and high-J CS lines trace different components. The
low-temperature component, on average over the sources, has a
T$_{rot}<$ 30K, N(CS)$> 1.0\times 10^{14}$cm$^{-2}$ and
n(H$_{2}$)$< 1.6\times 10^{5}$cm$^{-3}$, while the
high-temperature component has a T$_{rot}>$ 45K with N(CS)$<
1.0\times 10^{14}$cm$^{-2}$ and n(H$_{2}$)$> 2.5\times
10^{5}$cm$^{-3}$. For the brightest sources, we compared the
physical properties of these two CS components with the physical
properties derived from other tracers of dense gas, when
available. For the M~82-NE and M~82-SW molecular lobes, the
low-temperature CS component is in very good agreement with the
low-temperature component of methanol (see \citealt{Mart06b}).
Indeed their predicted (from LVG) kinetic temperatures are in
agreement within a factor of $\approx$2-3 while the gas densities
and the column densities are in agreement within a factor of $<$ 3
and $<$ 2, respectively. Similarly, the physical properties of the
high-temperature component of the CS is in good agreement with
those of the high-temperature component of the methanol. Hence, it
may be that the warm and dense methanol gas (respectively cold and
less dense methanol gas) might be well mixed with the warm and
dense CS gas (respectively cold and less dense CS gas) in these
two positions. We do not know if this correlation is common to all
PDR-dominated galaxies, or if it depends on the galaxy type.
Unfortunately, in the literature, no sample of methanol lines as
complete as the one provided by \citet{Mart06b} for M~82 exists.
Detections of methanol in another starburst-dominated galaxy,
NGC~253, exist (see e.g. \citealt{Mart06a}) but the authors do not
distinguish between NGC253-1 and NGC253-2 and do not provide LVG
predictions. In the same vein, in pure AGN-dominated galaxies
(such as NGC~1068) or in merging-dominated systems (such as the
Antennae Galaxies), no large sample of methanol lines have been
analyzed under LVG approximations so far.

Similarly to the molecular lobes of M~82, in both NGC253-1 and
NGC253-2, the low-temperature CS component show physical
properties in very good agreement with the properties derived for
HCN (single-temperature component whose properties are derived
using an LTE analysis; see \citealt{Mart06a}). The rotational
temperature for HCN and CS in NGC253-1 (and NGC 253-2) are in
agreement within a factor of $<$2 (respectively $<$ 1.4) while the
HCN and the CS column densities in NGC253-1 (and NGC 253-2) agree
within a factor of $<$ 1.5 (respectively $<$ 2.9). Similarly, in
both NGC 253 sources, the high-temperature CS component show
physical properties in agreement with those derived from
HCO$^{+}$, H$_{2}$CO and HNC (single-temperature component whose
properties are derived using an LTE analysis; see
\citealt{Mart06a}). This shows that the HCN and the
low-temperature CS gas on one side, and the HCO$^{+}$, H$_{2}$CO,
HNC and the high-temperature CS gas on the other side form two
distinguished groups.

Furthermore, it is of particular interest to compare CS and HCN
data in NGC~1068 (AGN-dominated galaxy). For this source,
\citet{Krip08} presented a low-temperature HCN component from LVG
modelling constrained by a reasonable sample of HCN lines. This
HCN gas component has T$_{K}$, n(H$_{2}$), N(CS) = 20K,
$10^{4}-10^{4.5}$ cm$^{-3}$, $10^{15}-10^{16}$ cm$^{-2}$ which is
a less dense and higher column density component than either our
low-temperature or our high-temperature CS component. However, 
taking into account the limitations and uncertainties of LVG models, 
it may not be appropriate to compare the two sets of data.

In parallel, for M~82-NE and M~82-SW molecular lobes, Table
\ref{tab:obs} shows a significant narrowing of the FWHM of the
CS(7-6) lines, as compared with the FWHM of the lower-J CS
transitions. This feature has already been observed and analyzed
by \citet{Baye08b} in the nucleus position. They interpreted this
characteristic as the presence of multiple gas components in the
source, the CS(7-6) line tracing a gas component warmer and denser
than the gas traced by lower-J CS lines. Population rotational
diagrams as well as LVG analysis (see Subsects. \ref{subsec:lte}
and \ref{subsec:lvg}) confirm that this feature also appears in
the molecular lobes of M~82. However such narrowing does not
appear in the other sources. Observations of the CS(7-6)
transition in a larger sample of galaxies are needed in order to
reveal very dense gas properties in PDR-dominated galaxies.

\section{Conclusions}\label{sec:con}

In order to better determine the properties of the very dense
star-forming gas over a large range of physical conditions, we
present the most complete, extragalactic CS survey ever performed
on nearby galaxies. In particular, we detect for the first time
the CS(5-4) and CS(7-6) lines, in various environments such as
starburst, AGN-dominated galaxies, irregular galaxies, merging
galaxies, etc.

In this first paper out of a series, we simply analyzed the data
through a rotational diagram method and an LVG modelling and we
show that for most sources at least two gas components are needed
to reproduce the observations. The low-temperature gas component
properties, derived from the CS(2-1), CS(3-2) and CS(4-3) lines
(low-J CS lines), vary from source to source within the range
T$_{K}$, n(H$_{2}$), N(CS) = 10-30K, 0.16-1.60$\times
10^{5}$cm$^{-3}$, 0.25-6.30$\times 10^{14}$cm$^{-2}$,
respectively. The high-temperature gas component properties,
derived from the CS(5-4) and CS(7-6) lines (high-J CS lines), vary
from source to source within the range T$_{K}$, n(H$_{2}$), N(CS)
= 45-70K, 6.30-40.0$\times 10^{5}$cm$^{-3}$, 0.16-1.00$\times
10^{14}$cm$^{-2}$, respectively. We also show that, using the
current data set of CS observations, in Overlap and M~83, it
appears that the very dense gas may be more homogeneously
distributed than in other nearby sources.However, we again 
underline here that our estimates of the physical properties 
of these galaxies suffer from LVG degeneracy.

Comparison with other tracers of dense gas such as HCN, HNC,
HCO$^{+}$, etc even in the brightest sources of our sample is very
difficult to perform. However, in the molecular lobes of M~82, we
have been able to show a good correlation between the CS and the
methanol. In both NGC~253-1 and NGC~253-2, we showed that the HCN
and the low-temperature CS gas on one side, and the HCO$^{+}$,
H$_{2}$CO, HNC and the high-temperature CS gas on the other side
have compatible rotational temperatures and column densities.

In pure AGN-dominated galaxy such as NGC~1068, we find no
correlation between the HCN and CS.

Forthcoming papers will focuss on detailed studies of the very
dense gas properties in individual nearby sources, investigating
for instance their dense gas star-formation efficiencies, their
dense gas masses based on CS modelling, etc, as well as on a more
statistical analysis such as Kennicutt-Schmidt laws in both
extragalactic and galactic sources.

\begin{landscape}
\begin{table}
  \caption{Observational parameters and Gaussian fits for
  the data set of CS observations. Where data were not
  available, we have put a black dash in the corresponding
  place in the table.}\label{tab:obs}
  \resizebox{18.8cm}{!}{
    \begin{tabular}{l c c c c c c c c c c c c c}
   \hline
   Source & Line & $\nu$ & Tsys & beam & Telesc.
          & $\int$(T$_{mb}$ dv) & V$_{peak}$ & FWHM & T$_{peak}$ & rms
          & RA(J2000) & DEC(J2000) & Refs$^{1}$\\
          & & & & size & name & & & & & & & & \\
          & & (GHz) & (K) & ($''$) &  &  (Kkms$^{-1}$)
          & (kms$^{-1}$) & (kms$^{-1}$) & (mK) & (mK)
          & (h:m:s) & ($^{\circ}$ : ' : '') &\\
   \hline
   NGC 253-1  & CS(2-1) & 97.980 & - & 51.0 & SEST
             & 4.4$\pm$0.5 & 172 & 96 & 43.4
             & 5.2 & 00:47:33.4 & -25:17:23.0 & d\\
             & CS(3-2) & 146.969 & - & 16.7 & IRAM-30m
             & 11.9$\pm$0.2 & 185 & 100 & 111.2
             & 4.1 & '' & '' & d\\
             & CS(4-3) & 195.954 & - & 12.6 & IRAM-30m
             & 11.5$\pm$0.4 & 185 & 100 & 108.2 & 7.0
             & '' & '' & d\\
             & CS(5-4) & 244.936 & - & 21.0 & SEST & 4.4$\pm$0.4 & 158
             & 107 & 38.2 & 10.8 & '' & '' & d\\
             %& CS(7-6) & 342.883 & 1398 & 14.0 & JCMT  & 16.5$\pm$2.9 & 211.9$\pm$2.2
             %& 43.8$\pm$4.2 & 353.4 & 30.0 & '' & '' & a\\
             & CS(7-6) & 342.883 & 679 & 14.0 & JCMT  & 7.9$\pm$0.3 & 185$^{2}$
             & 110.1$\pm$4.4 & 67.1 & 5.1 & '' & '' & a\\
   \hline
   NGC 253-2  & CS(2-1) & 97.980 & - & 51.0 & SEST
             & 8.2$\pm$0.5 & 290 & 116 & 66.5
             & 5.2 & 00:47:33.4 & -25:17:23.0 & d\\
             & CS(3-2) & 146.969 & - & 16.7 & IRAM-30m
             & 13.7$\pm$0.2 & 288 & 117 & 110.3
             & 4.1
             & '' & '' & d\\
             & CS(4-3) & 195.954 & - & 12.6 & IRAM-30m
             & 13.4$\pm$0.5 & 288 & 121 & 104.5
             & 7.0 & '' & '' & d\\
             & CS(5-4) & 244.936 & - & 21.0 & SEST  & 5.4$\pm$1.3 & 262
             & 108 & 46.7 & 10.8 & '' & '' & d\\
             %& CS(7-6) & 342.883 & 1398 & 14.0 & JCMT & 14.2$\pm$3.1 & 267.6$\pm$8.5
             %& 75.1$\pm$13.0 & 177.7 & 30.0 & '' & '' & a\\
             & CS(7-6) & 342.883 & 679 & 14.0 & JCMT  & 4.4$\pm$0.3 & 288$^{2}$
             & 107.1$\pm$7.1 & 38.7 & 5.1 & '' & '' & a\\
   \hline
   NGC 1068  & CS(3-2) & 146.969 & 600 & 16.7 & IRAM-30m & 9.1$\pm$1.5 & 1100
             & 245.0 & 30.0 & 10.0 & 02:42:40.7 & -00:00:47.6 & e\\
             & CS(5-4) & 244.936 & - & 10.0 & IRAM-30m & 3.3$\pm$0.3 & 1125$\pm$9.0 & 180.0$\pm$20.0
             & 17.6 & 3.0 & '' & '' & f\\
             & CS(7-6) & 342.883 & 307 & 14.0 & JCMT  & 1.4$\pm$0.5 & 1125.4$\pm$49.0
             & 279.0$\pm$99.3 & 4.8 & 8.1 & '' & '' & a\\
   \hline
   IC 342    & CS(1-0) & 48.991 & 350-550 & 36.0 & NRO
             & 2.5$\pm$0.4 & 30 & 50-60 & 50.0
             & 12.0 & 03:46:48.3 & 68:05:46.0 & g\\
             & CS(2-1) & 97.980 & 188 & 25.1 & IRAM-30m
             & 5.0$\pm$0.1 & 31.7$\pm$0.6 & 53.9$\pm$1.5 & 87.4
             & 2.5 & '' & '' & c\\
             & CS(3-2) & 146.969 & 225 & 16.7 & IRAM-30m
             & 4.9$\pm$0.1 & 31.2$\pm$0.6 & 51.7$\pm$1.6 & 88.6
             & 3.0 & '' & '' & c\\
             & CS(5-4) & 244.936 & 378 & 10.0 & IRAM-30m  & 2.5$\pm$0.2 & 38.2$\pm$2.0 & 63.4$\pm$5.1
             & 37.6 & 2.9 & '' & '' & c\\
             & CS(7-6) & 342.883 & 525 & 14.0 & JCMT  & 1.0$\pm$0.4 & 29.3$\pm$15.3
             & 75.6$\pm$30.6 & 12.3 & 8.5 & '' & '' & a\\
   \hline
   Henize 2-10 & CS(7-6) & 342.883 & 252 & 14.0 & JCMT  & 0.6$\pm$0.2 & 883.1$\pm$7.4
             & 42.0$\pm$16.1 & 13.0 & 7.4 & 08:36:15.2 & -26:24:34.0 & a\\
   \hline
   M~82\_NE  & CS(2-1) & 97.980 & 236 & 25.1 & IRAM-30m
             & 9.4$\pm$0.2 & 299.9$\pm$1.3 & 103.4$\pm$3.3 & 85.7
             & 8.5 & 09:55:54.4 & 69:40:54.6 & a\\
             & CS(3-2) & 146.969 & 278 & 16.7 & IRAM-30m
             & 8.9$\pm$0.1 & 290.9$\pm$0.7 & 114.6$\pm$2.0 & 73.3
             & 1.7 & '' & '' & a\\
             & CS(4-3) & 195.954 & 1425 & 12.6 & IRAM-30m
             & 6.5$\pm$0.8 & 309.2$\pm$6.1 & 87.7$\pm$12.1 & 70.1
             & 17.2 & '' & '' & a\\
             & CS(5-4) & 244.936 & 772 & 10.0 & IRAM-30m  & 6.4$\pm$0.2 & 313.0$\pm$1.1 & 73.5$\pm$2.6
             & 82.1 & 4.8 & '' & '' & a\\
             %& C$^{34}$S(5-4) & 241.016 & 1920 & 10.2 & IRAM-30m & 163
             %& 10.0 & - & - & - & - & 10.7 & 0.52 & 09:55:51.9 & 69:40:47.1 & a\\
             & CS(7-6) & 342.883 & 309 & 14.0 & JCMT  & 1.1$\pm$0.2 & 313.2$\pm$4.4
             & 58.1$\pm$6.8 & 15.7 & 2.9 & '' & '' & a\\
   \hline
    M~82\_SW & CS(2-1) & 97.980 & 237 & 25.1 & IRAM-30m
             & 8.9$\pm$0.2 & 131.4$\pm$1.2 & 109.8$\pm$2.7 & 75.9
             & 3.8 & 09:55:49.4 & 69:40:39.6 & a\\
             & CS(4-3) & 195.954 & 1628 & 12.6 & IRAM-30m
             & 7.3$\pm$1.0 & 140.0$\pm$6.6 & 104.7$\pm$17.2 & 65.5
             & 17.9
             & '' & '' & a\\
             %& C$^{34}$S(5-4) & 241.016 & 2945 & 10.2 & IRAM-30m & 29
             %& 10.0 & - & - & - & - & 39.4 & 0.52 & 09:55:51.9 & 69:40:47.1 & a\\
             & CS(7-6) & 342.883 & 340 & 14.0 & JCMT & 1.0$\pm$0.2 & 110.2$\pm$8.8
             & 60.3$\pm$14.8 & 13.8 & 5.1 & '' & '' & a\\
   \hline
   NGC 4038  & CS(2-1) & 97.980 & 201 & 25.1 & IRAM-30m
             & 1.0$\pm$0.1 & 1646.8$\pm$3.0 & 79.6$\pm$6.7 & 11.6
             & 2.2 &  12:01:52.8 & -18:52:05.3 & a\\
             & CS(3-2) & 146.969 & 471 & 16.7 & IRAM-30m
             & 0.8$\pm$0.1 & 1645.2$\pm$3.8 & 62.2$\pm$7.6 & 11.4
             & 2.6
             & '' & '' & a\\
             & CS(4-3) & 195.954 & 1396 & 12.6 & IRAM-30m
             & 1.1$\pm$0.3 & 1646.0$\pm$8.6 & 65.4$\pm$15.8 & 15.3
             & 6.5
             & '' & '' & a\\
             & CS(5-4) & 244.936 & 326 & 20.0 & JCMT  & 1.6$\pm$0.2 & 1649.9$\pm$6.5
             & 103.2$\pm$12.7 & 14.4 & 6.4  & '' & '' & b\\
             & CS(5-4) & 244.936 & 1924 & 10.5 & IRAM-30m  & 1.9$\pm$0.4 & 1666.3$\pm$9.7
             & 103.0 & 17.1 & 9.5 & '' & '' & a\\
             & CS(7-6) & 342.883 & 250 & 14.0 & JCMT  & 0.7$\pm$0.1 & 1629.4$\pm$7.6
             & 88.6$\pm$15.0 & 7.8 & 2.7 & '' & '' & a\\
   \hline
   NGC 4039  & CS(2-1) & 97.980 & 226 & 25.1 & IRAM-30m
             & 1.7$\pm$0.2 & 1640.8$\pm$16.3 & 228.2$\pm$33.6 & 7.1
             & 3.2 & 12:01:53.5 & -18:53:11.3 & a\\
             & CS(3-2) & 146.969 & 547 & 16.7 & IRAM-30m
             & 1.3$\pm$0.3 & 1691.6$\pm$19.1 & 145.0$\pm$30.2 & 8.3
             & 5.8
             & '' & '' & a\\
             & CS(4-3) & 195.954 & 1275 & 12.6 & IRAM-30m
             & - & - & - & -
             & 17.9
             & '' & '' & a\\
             & CS(5-4) & 244.936 & 277 & 20.0 & JCMT  & 1.4$\pm$0.2 & 1757.6$\pm$12.7
             & 157.1$\pm$26.5 & 8.2 & 4.0 & '' & '' & a\\
             %& CS(7-6) & 342.883 & 328 & 14.0 & JCMT & 0.4$\pm$0.1 & 1823.2$\pm$25.4
             %& 122.4$\pm$55.04 & 3.0 & 3.6 & '' & '' & a\\
   \hline
   OVERLAP   & CS(2-1) & 97.980 & 203 & 25.1 & IRAM-30m
             & 0.9$\pm$0.2 & 1498.8$\pm$5.6 & 81.2$\pm$18.9 & 10.6
             & 2.7
             & 12:01:54.9 & -18:52:59.0 & a\\
             & CS(3-2) & 146.969 & 467 & 16.7 & IRAM-30m
             & 0.8$\pm$0.2 & 1504.8$\pm$6.4 & 61.8$\pm$18.7 & 11.7
             & 4.8
             & '' & '' & a\\
             & CS(4-3) & 195.954 & 1290 & 12.6 & IRAM-30m
             & - & - & - & -
             & 14.0
             & '' & '' & a\\
             & CS(5-4) & 244.936 & 1763 & 10.0 & IRAM-30m & - & -
             & - & - & 14.4 & '' & '' & a\\
             & CS(7-6) & 342.883 & 230 & 14.0 & JCMT  & - & -
             & - & - & 4.3 & '' & '' & a\\
   \hline
   M~83      & CS(3-2) & 146.969 & 600 & 16.7 & IRAM-30m & 1.2$\pm$0.3 & 553.0$\pm$11.0
             & 78.0$\pm$24.0 & 15.0 & - & 13:36:59.2 & -29:52:04.5 & e\\
             & CS(5-4) & 244.936 & - & 10.0 & IRAM-30m  & 2.3$\pm$0.3 & 504.0$\pm$7.0
             & 82.4$\pm$1.1 & 26.5 & 0.5 & '' & '' & d\\
             & CS(7-6) & 342.883 & 121 & 14.0 & JCMT  & 0.4$\pm$0.1 & 540.0$\pm$9.5
             & 99.1$\pm$18.8 & 3.3 & 1.1 & '' & '' & a\\
  \hline
  \end{tabular}}

$^{1}$ Refs: a: This work, b: See \citet{Baye08b}, c: See
\citet{Alad09}, d: \citet{Mart05}, e: \citet{Maue89b}, f:
\citet{Mart09} and g: \citet{Pagl95}; $^{2}$: The velocity
positions of the two components in NGC 253 have been fixed to 185
kms$^{-1}$ and 288 kms$^{-1}$, consistently to what it has been
performed in \citet{Mart05}.
\end{table}
\end{landscape}

\begin{table*}
\caption{Results of the linear regressions (slope, correlation
coefficient, total source-averaged CS column density and
rotational temperature) for a single- and two-components fit. The
linear regressions have been obtained using the xmgrace software,
including in the calculations the error bars of the
observations.}\label{tab:T_rot}
\begin{tabular}{ccccc}
\hline
Source (Nb. of & slope & Correl. & N(CS)$^{1}$ &  T$_{rot}$ $^{1}$\\
fit components) & ($\times 10^{-2}$) & Coeff. & ($\times 10^{14}$) & (in K)\\
\hline
NGC 253-1 (one) & -2.67 & 0.878 & 1.79 & 16.3\\
NGC 253-1 (two) & -7.14 & 0.970 & 2.65 & 6.1 \\
                & -1.33 & 0.969 & 0.80 & 32.7 \\
\hline
NGC 253-2 (one) & -3.40 & 0.910 & 2.47 & 12.8\\
NGC 253-2 (two) & -8.34 & 0.952 & 4.39 & 5.2\\
                & -2.14 & 0.999 & 0.97 & 20.3 \\
\hline
NGC 1068 (one)  & -3.13 & 0.921 & 2.88 & 13.9\\
NGC 1068 (two)  & -6.09 & 1 & 5.80 & 7.1\\
                & -1.31 & 1 & 0.68 & 33.2\\
\hline
IC 342 (one) & -4.26 & 0.936 & 2.67 & 10.2\\
IC 342 (two) & -10.87 & 0.997 & 2.87 & 4.0 \\
             & -1.50 & 1 & 0.27 & 28.9 \\
\hline
M82-NE (one) & -3.91 & 0.958 & 2.64 & 11.1\\
M82-NE (two) & -7.81 & 0.992 & 4.25 & 5.6\\
             & -2.77 & 0.999 & 1.04 & 15.7\\
\hline
M82-SW (one) & -3.94 & 0.975 & 1.80 & 11.0\\
M82-SW (two) & -7.10 & 1 & 2.39 & 6.1\\
             & -3.02 & 1 & 0.74 & 14.4\\
\hline
size=7 $''$ & & & & \\
NGC 4038 (one) & -2.38 & 0.895 & 0.25 & 18.3 \\
NGC 4038 (two) & -6.24 & 0.936 & 0.29 & 7.0\\
               & -1.52 & 0.991 & 0.15 & 28.6\\
size=25\%$\times$7 $''$ & & & & \\
NGC 4038 (one) & -2.43 & 0.891 & 0.42 & 17.9 \\
NGC 4038 (two) & -6.49 & 0.943 & 0.51 & 6.7\\
               & -1.48 & 0.993 & 0.24 & 29.2\\
size=50\%$\times$7 $''$ & & & & \\
NGC 4038 (one) & -2.46 & 0.887 & 0.88 & 17.6 \\
NGC 4038 (two) & -6.67 & 0.947 & 1.10 & 6.5\\
               & -1.45 & 0.995 & 0.49 & 29.9\\
size=75\%$\times$7 $''$ & & & & \\
NGC 4038 (one) & -2.47 & 0.885 & 0.34 & 17.6 \\
NGC 4038 (two) & -6.74 & 0.949 & 4.29 & 6.5\\
               & -1.42 & 0.997 & 1.85 & 30.5\\
variable size & & & & \\
NGC 4038 (one) & -1.90 & 0.885 & 0.36 & 22.9 \\
NGC 4038 (two) & -5.40 & 0.962 & 0.34 & 8.0\\
               & -1.16 & 0.928 & 0.26 & 37.5\\
\hline
NGC 4039 (one) & -3.21 & 0.847 & 0.81 & 13.5 \\
NGC 4039 (two) & -11.34 & 1 & 1.01 & 3.8 \\
               & -1.27  & 1 & 0.55 & 34.1 \\
\hline
Overlap (one) & -5.67 & 1 & 0.07 & 7.7\\
\hline
M 83 (one)    & -2.73 & 0.999 & 0.45 & 15.9\\
\hline
\end{tabular}

 $^{1}$: The total source-averaged CS column
 density is derived using the following formulae:
 N(CS)=$10^{intersect} \times$Q$_{rot}$(T$_{rot}$)
 where \textit{intersect} is the value of the intersection
 of the linear regression with the y-axis (see Figs. \ref{fig:11}
 to \ref{fig:16}) and Q$_{rot}$(T$_{rot}$) is the partition
 function at the rotational temperature T$_{rot}$ (linear
 interpolation realized when the available range of T$_{rot}$ was not
 relevant for the studied case - see text). The rotational
 temperature is calculated using the formulae: T$_{rot} =
 - \frac{1}{slope} \times log(e)$.
\end{table*}

\begin{table}
\caption{Results of the LVG model analysis: the physical
properties of the best LVG model (having the lowest-$\chi^{2}$
value) for both the low- and high-temperature gas components. We
remind the reader that these values are only indicative since
degeneracy in LVG models occurs.}\label{tab:LVG}
\begin{center}
\begin{tabular}{ccccc}
\hline
Source & T$_{K}$ & n(H$_{2}$) & N(CS) \\
 & in K &  ($\times 10^{5}$ cm$^{-3}$) & ($\times 10^{14}$cm$^{-2}$) &\\
\hline
NGC 253-1 & 30 & 0.40 & 2.50 \\
          & 65 & 25.0 & 0.63 \\
\hline
NGC 253-2 & 15 & 0.16 & 4.00\\
          & 70 & 10.0 & 1.00\\
\hline
NGC 1068 & 20 & 1.50 & 6.30\\
         & 65 & 40.0 & 0.63\\
\hline
IC 342   & 15 & 0.63 & 1.00 \\
         & 50 & 16.0 & 0.25 \\
\hline
M82-NE   & 10 & 1.00 & 4.00 \\
         & 65 & 6.30 & 1.00 \\
\hline
M82-SW  & 15 & 1.60 & 2.50 \\
        & 45 & 6.30 & 0.63\\
\hline
NGC 4038 & 10 & 1.00 & 0.25 \\
         & 50 & 40.0 & 0.16 \\
\hline
NGC 4039 & 10 & 0.40 & 1.00 \\
         & 45 & 2.50 & 0.40 \\
\hline
Overlap  & 10 & 2.50 & 0.06 \\
\hline
M 83     & 65 & 6.30 & 0.15 \\
\hline
\end{tabular}
\end{center}
\end{table}

\section*{Acknowledgments}

EB acknowledges financial support from the Leverhulme Trust and M.
Banerji for her participation to one the observational sessions on
the top of the Mauna Kea. The James Clerk Maxwell Telescope is
operated by The Joint Astronomy Centre on behalf of the Science
and Technology Facilities Council of the United Kingdom, the
Netherlands Organization for Scientific Research, and the National
Research Council of Canada. Authors acknowledge the anonymous
referee for his/her useful comments which significantly improved
the paper.

\bibliographystyle{apj}
\bibliography{references}

\begin{thebibliography}{35}
\expandafter\ifx\csname natexlab\endcsname\relax\def\natexlab#1{#1}\fi

\bibitem[{{Aladro} {et~al.}(2009){Aladro}, {Martin}, {Martin-Pintado}, \&
  {Bayet}}]{Alad09}
{Aladro}, R., {Martin}, S., {Martin-Pintado}, J., \& {Bayet}, E. 2009, \apj, in
  preparation

\bibitem[{{Bayet} {et~al.}(2004){Bayet}, {Gerin}, {Phillips}, \&
  {Contursi}}]{Baye04}
{Bayet}, E., {Gerin}, M., {Phillips}, T.~G., \& {Contursi}, A. 2004, \aap, 427,
  45

\bibitem[{{Bayet} {et~al.}(2006){Bayet}, {Gerin}, {Phillips}, \&
  {Contursi}}]{Baye06}
---. 2006, \aap, 460, 467

\bibitem[{{Bayet} {et~al.}(2008{\natexlab{a}}){Bayet}, {Lintott}, {Viti},
  {Mart{\'{\i}}n-Pintado}, {Mart{\'{\i}}n}, {Williams}, \&
  {Rawlings}}]{Baye08b}
{Bayet}, E., {Lintott}, C., {Viti}, S., {Mart{\'{\i}}n-Pintado}, J.,
  {Mart{\'{\i}}n}, S., {Williams}, D.~A., \& {Rawlings}, J.~M.~C.
  2008{\natexlab{a}}, \apjl, 685, L35

\bibitem[{{Bayet} {et~al.}(2008{\natexlab{b}}){Bayet}, {Viti}, {Williams}, \&
  {Rawlings}}]{Baye08a}
{Bayet}, E., {Viti}, S., {Williams}, D.~A., \& {Rawlings}, J.~M.~C.
  2008{\natexlab{b}}, \apj, 676, 978

\bibitem[{{Bronfman} {et~al.}(1996){Bronfman}, {Nyman}, \& {May}}]{Bron96}
{Bronfman}, L., {Nyman}, L.-A., \& {May}, J. 1996, \aaps, 115, 81

\bibitem[{{de Jong} {et~al.}(1975){de Jong}, {Dalgarno}, \& {Chu}}]{DeJo75}
{de Jong}, T., {Dalgarno}, A., \& {Chu}, S.-I. 1975, \apj, 199, 69

\bibitem[{{Garc{\'{\i}}a-Burillo} {et~al.}(2001){Garc{\'{\i}}a-Burillo},
  {Mart{\'{\i}}n-Pintado}, {Fuente}, \& {Neri}}]{Garc01}
{Garc{\'{\i}}a-Burillo}, S., {Mart{\'{\i}}n-Pintado}, J., {Fuente}, A., \&
  {Neri}, R. 2001, \apjl, 563, L27

\bibitem[{{Garc{\'{\i}}a-Burillo} {et~al.}(2002){Garc{\'{\i}}a-Burillo},
  {Mart{\'{\i}}n-Pintado}, {Fuente}, {Usero}, \& {Neri}}]{Garc02}
{Garc{\'{\i}}a-Burillo}, S., {Mart{\'{\i}}n-Pintado}, J., {Fuente}, A.,
  {Usero}, A., \& {Neri}, R. 2002, \apjl, 575, L55

\bibitem[{{Girart} {et~al.}(2002){Girart}, {Viti}, {Williams}, {Estalella}, \&
  {Ho}}]{Gira02}
{Girart}, J.~M., {Viti}, S., {Williams}, D.~A., {Estalella}, R., \& {Ho},
  P.~T.~P. 2002, \aap, 388, 1004

\bibitem[{{Goldreich} \& {Kwan}(1974)}]{Gold74}
{Goldreich}, P. \& {Kwan}, J. 1974, \apj, 189, 441

\bibitem[{{Goldsmith} \& {Langer}(1999)}]{Gold99}
{Goldsmith}, P.~F. \& {Langer}, W.~D. 1999, \apj, 517, 209

\bibitem[{{Greve} {et~al.}(2009){Greve}, {Papadopoulos}, {Gao}, \&
  {Radford}}]{Grev09}
{Greve}, T.~R., {Papadopoulos}, P.~P., {Gao}, Y., \& {Radford}, S.~J.~E. 2009,
  \apj, 692, 1432

\bibitem[{{Helfer} \& {Blitz}(1995)}]{Helf95}
{Helfer}, T.~T. \& {Blitz}, L. 1995, \apj, 450, 90

\bibitem[{{Klein} {et~al.}(1983){Klein}, {Sandford}, \& {Whitaker}}]{Klei83}
{Klein}, R.~I., {Sandford}, II, M.~T., \& {Whitaker}, R.~W. 1983, \apjl, 271,
  L69

\bibitem[{{Krips} {et~al.}(2008){Krips}, {Neri}, {Garc{\'{\i}}a-Burillo},
  {Mart{\'{\i}}n}, {Combes}, {Graci{\'a}-Carpio}, \& {Eckart}}]{Krip08}
{Krips}, M., {Neri}, R., {Garc{\'{\i}}a-Burillo}, S., {Mart{\'{\i}}n}, S.,
  {Combes}, F., {Graci{\'a}-Carpio}, J., \& {Eckart}, A. 2008, \apj, 677, 262

\bibitem[{{Larosa}(1983)}]{LaRo83}
{Larosa}, T.~N. 1983, \apj, 274, 815

\bibitem[{{Lintott} \& {Viti}(2006)}]{Lint06}
{Lintott}, C. \& {Viti}, S. 2006, \apjl, 646, L37

\bibitem[{{Mart{\'{\i}}n} {et~al.}(2006{\natexlab{a}}){Mart{\'{\i}}n},
  {Mart{\'{\i}}n-Pintado}, \& {Mauersberger}}]{Mart06b}
{Mart{\'{\i}}n}, S., {Mart{\'{\i}}n-Pintado}, J., \& {Mauersberger}, R.
  2006{\natexlab{a}}, \aap, 450, L13

\bibitem[{{Mart{\'{\i}}n} {et~al.}(2009){Mart{\'{\i}}n},
  {Mart{\'{\i}}n-Pintado}, \& {Mauersberger}}]{Mart09}
---. 2009, \apj, 694, 610

\bibitem[{{Mart{\'{\i}}n} {et~al.}(2005){Mart{\'{\i}}n},
  {Mart{\'{\i}}n-Pintado}, {Mauersberger}, {Henkel}, \&
  {Garc{\'{\i}}a-Burillo}}]{Mart05}
{Mart{\'{\i}}n}, S., {Mart{\'{\i}}n-Pintado}, J., {Mauersberger}, R., {Henkel},
  C., \& {Garc{\'{\i}}a-Burillo}, S. 2005, \apj, 620, 210

\bibitem[{{Mart{\'{\i}}n} {et~al.}(2003){Mart{\'{\i}}n}, {Mauersberger},
  {Mart{\'{\i}}n-Pintado}, {Garc{\'{\i}}a-Burillo}, \& {Henkel}}]{Mart03}
{Mart{\'{\i}}n}, S., {Mauersberger}, R., {Mart{\'{\i}}n-Pintado}, J.,
  {Garc{\'{\i}}a-Burillo}, S., \& {Henkel}, C. 2003, \aap, 411, L465

\bibitem[{{Mart{\'{\i}}n} {et~al.}(2006{\natexlab{b}}){Mart{\'{\i}}n},
  {Mauersberger}, {Mart{\'{\i}}n-Pintado}, {Henkel}, \&
  {Garc{\'{\i}}a-Burillo}}]{Mart06a}
{Mart{\'{\i}}n}, S., {Mauersberger}, R., {Mart{\'{\i}}n-Pintado}, J., {Henkel},
  C., \& {Garc{\'{\i}}a-Burillo}, S. 2006{\natexlab{b}}, \apjs, 164, 450

\bibitem[{{Mauersberger} \& {Henkel}(1989)}]{Maue89a}
{Mauersberger}, R. \& {Henkel}, C. 1989, \aap, 223, 79

\bibitem[{{Mauersberger} {et~al.}(1989){Mauersberger}, {Henkel}, {Wilson}, \&
  {Harju}}]{Maue89b}
{Mauersberger}, R., {Henkel}, C., {Wilson}, T.~L., \& {Harju}, J. 1989, \aap,
  226, L5

\bibitem[{{Meier} {et~al.}(2001){Meier}, {Turner}, {Crosthwaite}, \&
  {Beck}}]{Meie01b}
{Meier}, D.~S., {Turner}, J.~L., {Crosthwaite}, L.~P., \& {Beck}, S.~C. 2001,
  \aj, 121, 740

\bibitem[{{Muraoka} {et~al.}(2009){Muraoka}, {Kohno}, {Tosaki}, {Kuno},
  {Nakanishi}, {Handa}, {Sorai}, {Ishizuki}, \& {Okuda}}]{Mura09}
{Muraoka}, K., {Kohno}, K., {Tosaki}, T., {Kuno}, N., {Nakanishi}, K., {Handa},
  T., {Sorai}, K., {Ishizuki}, S., \& {Okuda}, T. 2009, \pasj, 61, 163

\bibitem[{{Paglione} {et~al.}(1995){Paglione}, {Jackson}, {Ishizuki}, \&
  {Rieu}}]{Pagl95}
{Paglione}, T.~A.~D., {Jackson}, J.~M., {Ishizuki}, S., \& {Rieu}, N. 1995,
  \aj, 109, 1716

\bibitem[{{Peng} {et~al.}(1996){Peng}, {Zhou}, {Whiteoak}, {Lo}, \&
  {Sutton}}]{Peng96}
{Peng}, R., {Zhou}, S., {Whiteoak}, J.~B., {Lo}, K.~Y., \& {Sutton}, E.~C.
  1996, \apj, 470, 821

\bibitem[{{Planesas} {et~al.}(1991){Planesas}, {Scoville}, \& {Myers}}]{Plan91}
{Planesas}, P., {Scoville}, N., \& {Myers}, S.~T. 1991, \apj, 369, 364

\bibitem[{{Plume} {et~al.}(1992){Plume}, {Jaffe}, \& {Evans}}]{Plum92}
{Plume}, R., {Jaffe}, D.~T., \& {Evans}, II, N.~J. 1992, \apjs, 78, 505

\bibitem[{{Schinnerer} {et~al.}(2008){Schinnerer}, {B{\"o}ker}, {Meier}, \&
  {Calzetti}}]{Schi08}
{Schinnerer}, E., {B{\"o}ker}, T., {Meier}, D.~S., \& {Calzetti}, D. 2008,
  \apjl, 684, L21

\bibitem[{{Turner}(1991)}]{Turn91}
{Turner}, B.~E. 1991, \apjs, 76, 617

\bibitem[{{Walker} {et~al.}(1990){Walker}, {Walker}, {Carlstrom}, \&
  {Martin}}]{Walk90}
{Walker}, C.~E., {Walker}, C.~K., {Carlstrom}, J.~E., \& {Martin}, R.~N. 1990,
  NASA Conference Publication, 3084, 78

\bibitem[{{Wilson} {et~al.}(2000){Wilson}, {Scoville}, {Madden}, \&
  {Charmandaris}}]{Wils00}
{Wilson}, C.~D., {Scoville}, N., {Madden}, S.~C., \& {Charmandaris}, V. 2000,
  \apj, 542, 120

\end{thebibliography}

\end{document}